\documentclass[envcountsame,envcountsect,orivec,12pt,%
 nolinenumbers,undated]{lnthi}

\usepackage{graphics}

\title{Fast Breadth-First Search in Still Less Space}

\author{Torben Hagerup}

\institute{\Tinfuna[5]\\
  \email{hagerup@informatik.uni-augsburg.de}}

\newenvironment{itemizewithlabel}
 {\begin{itemize}\def\makelabel##1{\hbox to 6pt{\hss\llap{##1}}}}
 {\end{itemize}}
\def\Tmu{\mu}

\begin{document}

\maketitle{}

\begin{abstract}
It is shown that a breadth-first search
in a directed or undirected graph with $n$ vertices
and $m$ edges can be carried out in $O(n+m)$ time with
$n\log_2 3+O((\log n)^2)$ bits of working memory.

\bigskip

\end{abstract}

\pagestyle{plain}
\thispagestyle{plain}

\section{Introduction}
\label{sec:intro}

\subsection{Space-Bounded Computation}

The study of the amount of memory necessary
to solve specific computational problems
has a long tradition.
A fundamental early result in the area
is the discovery by Savitch~\cite{Sav70} that the
$s$-$t$ connectivity problem
(given a graph $G$ and two vertices $s$ and $t$ in $G$,
decide whether
$G$ contains a path from $s$ to $t$)
can be solved
with $O((\log n)^2)$ bits of memory
on $n$-vertex graphs.
In order for this and related results to make sense,
one must distinguish between the memory used
to hold the input and the working memory,
which is the only memory accounted for.
The working memory is usable without
restrictions, but the memory that holds the
input is read-only
and any output is stored in write-only memory.
Informally, these conventions serve to forbid
``cheating'' by using input or output memory
for temporary storage.
They are all the more natural when, as in the
original setting of Savitch, the input graph is
present only in the form of a computational procedure
that can test the existence of an
edge between two given vertices.

Savitch's algorithm
is admirably frugal as concerns memory, but its
(worst-case) running time is superpolynomial.
It was later generalized by Barnes, Buss, Ruzzo
and Schieber~\cite{BarBRS98}, who proved,
in particular, that the $s$-$t$ connectivity
problem can be solved
on $n$-vertex graphs
in $n^{O(1)}$ time
using $O({n/{2^{b\sqrt{\log n}}}})$ bits
for arbitrary fixed~$b>0$.
In the special case of undirected graphs, a
celebrated result of Reingold~\cite{Rei08}
even achieves polynomial time
with just $O(\log n)$ bits.
The running times of the algorithms
behind the latter results,
although polynomial, are ``barely so''
in the sense that the polynomials are of high degree.
A more recent research direction searches for
algorithms that still use memory as sparingly
as possible but are nonetheless
fast, ideally as fast as the best algorithms
that are not subject to space restrictions.
The quest to reduce space requirements and running time
simultaneously is motivated in practical terms
by the existence of small mobile or embedded
devices with little memory,
by memory hierarchies that allow smaller data
sets to be processed faster, and by situations
in which the input is too big to be stored locally
and must be accessed through query procedures
running on a remote server.
The Turing machine models running time on true
computers rather crudely, so the model of computation
underlying the newer research is the
random-access machine and, more specifically, the word RAM.

\subsection{The Breadth-First-Search Problem}

This paper continues an ongoing search
for the best time and space bounds for
carrying out a breadth-first search or BFS in a
directed or undirected graph.
Formally, we consider the BFS problem
to be that of
computing a shortest-path spanning forest
of an input graph $G=(V,E)$
consistent with a given permutation of $V$
in top-down order,
a somewhat tedious definition of which
can be found in~\cite{Hag18}.
Suffice it here to say that
if all vertices of the input graph $G=(V,E)$
are reachable from a designated start vertex~$s\in V$,
the task at hand essentially is to
output the vertices in $V$ in an order
of nondecreasing distance from $s$ in~$G$.
The BFS problem is important in itself, but has also
served as a yardstick with which to gauge the
strength of new algorithmic and
data-structuring ideas in the realm of
space-efficient computing.

In the following consider an input graph
$G=(V,E)$ and take $n=|V|$ and $m=|E|$.
The algorithms of Savitch~\cite{Sav70}
and of Barnes et al.~\cite{BarBRS98} are easily
adapted, within the time and space
bounds cited above, to compute the actual distance
from $s$ to $t$ ($\infty$ if $t$ is
not reachable from~$s$).
As a consequence, the BFS problem can be solved
on $n$-vertex graphs
with $O((\log n)^2)$ bits
or in $n^{O(1)}$ time with
${n/{2^{\Omega(\sqrt{\log n})}}}$ bits.
Every reasonably fast BFS algorithm
known to the author, however, can be characterized
by an integer constant $c\ge 2$, dynamically assigns
to each vertex in $V$ one of $c$ states
or \emph{colors}, and maintains the color
of each vertex explicitly or implicitly.
Let us call such an algorithm a
\emph{$c$-color} BFS algorithm.
E.g., the classic BFS algorithm marks each
vertex as visited or unvisited and stores
a subset of the visited vertices in a
FIFO queue, which makes it a 3-color algorithm:
The unvisited vertices are \emph{white},
the visited vertices in the FIFO queue are
\emph{gray}, and the remaining visited
vertices are \emph{black}.
Because the distribution of colors over the
vertices can be nearly arbitrary, a $c$-color
BFS algorithm with an $n$-vertex input graph
must spend at least $n\log_2 c$ bits
on storing the vertex colors.
The classic BFS algorithm uses much more space
since the FIFO queue may hold nearly $n$ 
vertices and occupy $\Theta(n\log n)$ bits.

Similarly as Dijkstra's algorithm can be
viewed as an abstract algorithm
turned into a concrete algorithm
by the choice of a particular priority-queue
data structure,
Elmasry, Hagerup and Kammer~\cite{ElmHK15}
described a simple abstract 4-color BFS
algorithm that uses
$O(n+m)$ time plus $O(n+m)$ calls of operations
in an appropriate data structure
that stores the vertex colors.
This allowed them to derive a first BFS algorithm
that works in $O(n+m)$ time with $O(n)$ bits.
Using the same abstract algorithm with a different data
structure, Banerjee, Chakraborty and Raman~\cite{BanCR16}
lowered the space bound to
$2 n+O({{n\log\log n}/{\log n}})$ bits.
Concurrently, Hagerup and Kammer~\cite{HagK16} obtained a
space bound of $n\log_2 3+O({n/{(\log n)^t}})$ bits,
for arbitrary fixed $t\ge 1$, by
stepping to a better so-called
\emph{choice-dictionary} data structure but, more
significantly,
by developing an abstract 3-color BFS algorithm to
work with it.
The algorithm uses the three colors
white, gray and black and,
for an undirected graph in which all vertices are
reachable from the start vertex $s$,
can be described via the code below.
No output is mentioned,
but a vertex can be output
when it is colored gray.

\goodbreak

\begin{tabbing}
\quad\=\quad\=\quad\=\quad\=\kill
\>Color all vertices white;\\
\>Color $s$ gray;\\
\>\textbf{while} some vertex is gray \textbf{do}\\
\>\>\textbf{for each} gray vertex $u$ \textbf{do}\\
\>\>\>\textbf{if} $u=s$ or $u$ has a black neighbor \textbf{then}\\
\>\>\>\>Color gray all white neighbors of~$u$;\\
\>\>\textbf{for each} gray vertex $u$ \textbf{do}\\
\>\>\>\textbf{if} $u$ has no white neighbor \textbf{then}\\
\>\>\>\>Color $u$ black;
\end{tabbing}

Roughly speaking, the white vertices have not yet been
encountered by the search, the black vertices are
completely done with, and the gray vertices form the layer
of currently active vertices at a common distance from~$s$.
The two inner loops of the algorithm iterate over
the gray vertices in order to replace them by their
white neighbors, which form the next gray layer.
Both iterations are \emph{dynamic} in the sense
that the set of gray vertices is changing while
it is being iterated over.
The first iteration colors additional vertices gray,
and we would prefer for these newly gray vertices
not to be enumerated by the iteration.
Satisfying this requirement is not easy for a
space-efficient algorithm, however, and therefore
the iteration instead tests each
enumerated vertex for being
``old''---exactly then does it equal $s$ or have a
black neighbor---and ignores the other gray vertices.
Similarly, the second iteration colors black
only those gray vertices that are no longer needed
as neighbors of white vertices---these
include all ``old'' gray vertices.
Even so, the choice dictionary must support dynamic iteration
suitably.
This represents the biggest challenge for a
space-efficient implementation of the
abstract algorithm.

For a directed graph, the changes are slight:
``black neighbor'' should be replaced by
``black inneighbor'', and each of the two
occurrences of ``white neighbor''
should be replaced by
``white outneighbor''.
If not all vertices are reachable from~$s$,
the code above, except for its first line,
must be wrapped
in a standard way in an outer loop that
steps $s$ through all vertices in a suitable order
and restarts
the BFS at every vertex found to still
be white when it is chosen as~$s$.
This leads to no additional complications
and will be ignored in the following.

\subsection{Recent Work and Our Contribution}

Starting with the algorithm of Hagerup and Kammer~\cite{HagK16},
all new BFS algorithms have space bounds of the
form $n\log_2 3+s(n)$ bits for some function $s$
with $s(n)=o(n)$.
In a practical setting the leading factor of
$\log_2 3$ is likely to matter more than the
exact form of~$s$, so that the progress since
the algorithm of Hagerup and Kammer
could be viewed as insignificant.
However, at least from a theoretical point of view
it is interesting to explore how much space
is needed beyond the seemingly unavoidable
$n\log_2 3$ bits required to store the vertex colors.
If a 3-color BFS algorithm uses $n\log_2 3+s(n)$
bits, we will therefore say that it works
with $s(n)$ \emph{extra} bits.
If its running time is $t(n,m)$, we may summarize
its resource requirements in the pair
$(t(n,m),s(n))$.
Adapting the notion of pareto dominance,
we say that an algorithm with the resource
pair $(t(n,m),s(n))$ \emph{dominates} an
algorithm with the resource pair
$(t'(n,m),s'(n))$ if
$t(n,m)=O(t'(n,m))$ and $s(n)=o(s'(n))$ or
$t(n,m)=o(t'(n,m))$ and $s(n)=O(s'(n))$.

Banerjee, Chakraborty, Raman and Satti~\cite{BanCRS18}
indicated a slew of 3-color BFS algorithms
with the following resource pairs:
$(n^{O(1)},o(n))$,
$(O(n m),O((\log n)^2))$,
$(O(m(\log n)^2),o(n))$ and
$(O(m\log n f(n)),O({n/{f(n)}}))$
for certain slowly growing
functions~$f$.
The first of these algorithms is dominated by
that of Barnes et al.~\cite{BarBRS98},
which also uses polynomial time but $o(n)$ bits
altogether, not just $o(n)$ extra bits.
The third algorithm of~\cite{BanCRS18} is
dominated by the algorithm of
Hagerup and Kammer~\cite{HagK16}, whose
resource pair is
$(O(n+m),O({n/{(\log n)^t}}))$
for arbitrary fixed $t\ge 1$.
Instantiating the 3-color abstract algorithm of~\cite{HagK16}
with a new choice dictionary, Hagerup~\cite{Hag18}
obtained an algorithm that has the resource
pair $(O(n\log n+m\log\log n),O((\log n)^2))$
and dominates the two remaining algorithms
of~\cite{BanCRS18}.
Another algorithm of~\cite{Hag18}
is faster but less space-efficient and
has the resource pair
$(O(n\log n+m),n^\epsilon)$ for arbitrary
fixed $\epsilon>0$.

Here we present a new data structure, designed
specifically to be used with the abstract
3-color BFS algorithm of~\cite{HagK16},
that leads to a concrete BFS algorithm
working in $O(n+m)$ time using
$n\log_2 3+O((\log n)^2)$ bits of
working memory.
The new algorithm combines the best time and
space bounds of all previous algorithms with
running-time bounds of $O(n m)$ or less,
and therefore
dominates all of them.
It is also simpler than
several previous algorithms.
We obtain a slightly more general
result by introducing a tradeoff parameter
$t\ge 1$:
The running time is $O((n+m)t)$, and the
space bound is $n\log_2 3+O({{(\log n)^2}/t}+\log n)$
bits.
If the degrees of the vertices
$1,\ldots,n$
of the input graph $G$
form a nondecreasing sequence or
if $G$ is approximately regular,
we achieve a running time of
$O((n+m)\log\log n)$ with just
$n\log_2 3+O(\log n)$ bits.

The technical contributions of the
present paper include:
\begin{itemize}
\item
A new representation of vertex colors
\item
A new approach to dynamic iteration
\item
A refined analysis of the abstract 3-color
BFS algorithm of~\cite{HagK16} and of a data
structure of
Dodis, P{\v a}tra{\c s}cu and Thorup~\cite{DodPT10}
for storing nonbinary arrays
\item
An amortized analysis of the new data structure.
\end{itemize}
Conversely, we draw on~\cite{HagK16}
not only for its abstract 3-color BFS algorithm,
but also for setting many of the basic concepts straight
and for a technical lemma.
Another crucial component is the
in-place chain technique of Katoh and Goto~\cite{KatG17},
as developed further in~\cite{Hag18,KamS18}.
The fundamental representation of $n$ colors
drawn from $\{0,1,2\}$ in close to $n\log_2 3$
bits so as to support efficient access to
individual colors,
also alluded to above,
is due to Dodis
et al.~\cite{DodPT10}.

\section{Preliminaries}

We do not need to be very specific about the
way in which the input graph $G=(V,E)$ is
presented to the algorithm.
With $n=|V|$ and $m=|E|$,
we assume that $n$ can be determined
in $O(n+m)$ time and that $V=\{1,\ldots,n\}$.
If $G$ is undirected, we also assume that
for each vertex $u\in V$, it is possible
to iterate over the neighbors of $u$ in
at most constant time plus time proportional
to their number.
If $G$ is directed, the assumption is the same,
but now the neighbors of $u$ include both
the inneighbors and the outneighbors of~$u$,
and inneighbors should (of course)
be distinguishable from outneighbors.

Our model of computation is a word RAM~\cite{AngV79,Hag98}
with a word length of $w\in\mathbb{N}$ bits, where we assume that $w$ is
large enough to allow all memory words in use to be addressed.
As part of ensuring this,
we assume that $w\ge\log_2 n$.
The word RAM has constant-time operations
for addition, subtraction and multiplication
modulo $2^w$, division with truncation
($(x,y)\mapsto\Tfloor{{x/y}}$ for $y>0$),
left shift modulo $2^w$
($(x,y)\mapsto (x\ll y)\bmod 2^w$,
where $x\ll y=x\cdot 2^y$),
right shift
($(x,y)\mapsto x\gg y=\Tfloor{{x/{2^y}}}$),
and bitwise Boolean operations
($\textsc{and}$, $\textsc{or}$ and $\textsc{xor}$
(exclusive or)).

\section{The Representation of the Vertex Colors}
\label{sec:rep}

This section develops a data structure for
storing a color drawn from the set
$\{\mbox{white},\mbox{gray},\mbox{black}\}$
for each of the $n$ vertices in
$V=\{1,\ldots,n\}$.
The data structure enables linear-time execution
of the abstract 3-color BFS algorithm
of~\cite{HagK16} and occupies
$n\log_2 3+O((\log n)^2)$ bits.
An inspection of the algorithm reveals that
the operations that must be supported
by the data structure are
reading and updating the colors of given
vertices---this by itself is easy---and
dynamic iteration over the set of gray vertices.
A main constraint for the latter operation
is that the iteration must happen
in time proportional to the number of
gray vertices, i.e., we must be able to
find the gray vertices efficiently.

Let us encode the color white as $1=\texttt{01}_2$,
gray as $0=\texttt{00}_2$ and
black as $2=\texttt{10}_2$.
In the following we will not distinguish between
a color and its corresponding integer or
2-bit string.
Take $\Lambda=\Tfloor{\log_2 n}$, $q=10\Lambda$
and $\lambda=\Tceil{\log_2 q}=\Theta(\log\log n)$.
In the interest of simplicity let us
assume that $n$ is large enough to make
$\lambda^2\le\Lambda$.
In order to keep track of the colors of the
vertices in $V$
we divide the sequence of $n$ colors into
$N=\Tfloor{n/q}$ \emph{segments}
of exactly $q$ colors each, with at most
$q-1$ colors left over.
Each segment is represented via a
\emph{big integer} drawn from
$\{0,\ldots,3^q-1\}$.
Because $q=O(\log n)$, big integers can be
manipulated in constant time.
The $N$ big integers are in turn maintained
in an instance of the data structure of
Lemma~\ref{lem:dodis} below, which occupies
$N\log_2 3^q+O((\log N)^2)=
n\log_2 3+O((\log n)^2)$ bits.

\begin{lemma}[\normalfont{\cite{DodPT10}, Theorem~1}\bf]
\label{lem:dodis}
There is a data structure that, given arbitrary
positive integers $N$ and $C$ with
$C=N^{O(1)}$, can be initialized
in $O(\log N)$ time and subsequently
maintains an array of $N$ elements drawn from
$\{0,\ldots,C-1\}$ in
$N\log_2 C+O((\log N)^2)$ bits
such that individual array elements
can be read and updated in constant time.
\end{lemma}

\subsection{Containers and Their Structure and Operations}

\begin{figure}
\begin{center}
\includegraphics{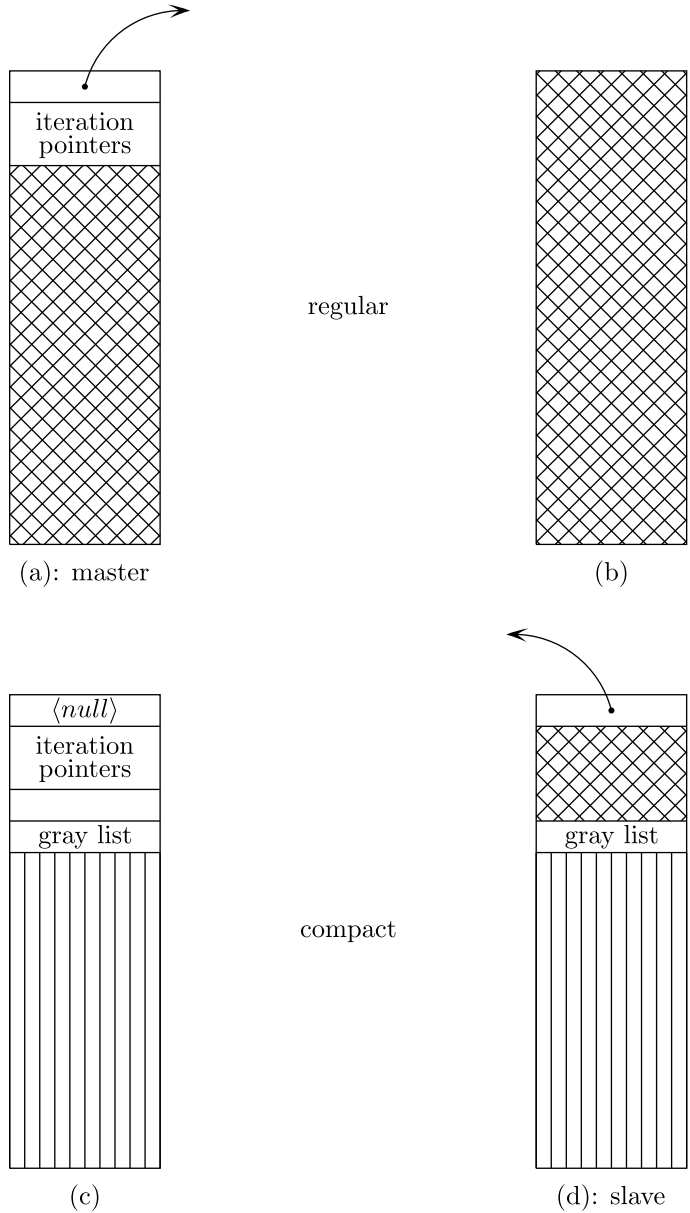}
\end{center}
\caption{Four different
representations in containers.
Crosshatched areas symbolize colors
white, gray and black
stored to base~3, while vertically striped
areas symbolize
black-and-white vectors.}
\label{fig:rep}
\end{figure}

We view the $N$ big integers as objects
with a nontrivial internal structure and
therefore use the more suggestive term
\emph{container} to denote the big integer
in a given position in the sequence of
$N$ big integers.
We shall say that each of the $q$ vertices
whose colors are stored in a container
is \emph{located}
in the container.
A container may represent $q$ colors
$a_0,\ldots,a_{q-1}$ in several different ways
illustrated in Fig.~\ref{fig:rep}.
The most natural representation
is as the integer
$x=\sum_{j=0}^{q-1}a_j 3^j$.
We call this the \emph{regular representation},
and a container is \emph{regular}
if it uses the regular representation
(Fig.~\ref{fig:rep}(b)).
When a vertex $u$ is located in a regular container,
we can read and update the color of~$u$
in constant time, provided
that we store a table
of the powers $3^0,3^1,\ldots,3^{q-1}$.
E.g., with notation as above,
$a_j=\Tfloor{{x/{3^j}}}\bmod 3$
for $j=0,\ldots,q-1$.
The table occupies $O((\log n)^2)$ bits and
can be computed in $O(\log n)$ time.

We allow a variant in which a regular
container $D$ is a \emph{master}
(Fig.~\ref{fig:rep}(a)).
The difference is that
the $3\Lambda$ most significant
bits of the big integer corresponding to~$D$
are relocated to a different container
$D'$, said to be the
\emph{slave} corresponding to $D$, and stored there.
This frees $3\Lambda-1$ bits in $D$ for other uses
(the most significant bit is fixed at~\texttt{0}
to ensure that the value of the big digit
does not exceed $3^q-1$).
Since $\Tceil{\log_2(N+1)}\le
\log_2({n/8})+1=\log_2 n-2
\le\Tfloor{\log_2 n}-1=\Lambda-1$,
we can store a pointer (possibly \emph{null})
to a container in $\Lambda-1$ bits,
so a master
has room for three such pointers.
One of these designates the slave~$D'$,
while the use of the two other pointers,
called \emph{iteration pointers},
is explained later.
Even though it may be necessary to access the
data relocated to the slave, a master
still allows vertex colors to be
read and updated in constant time.

When it is desired to iterate over
the gray vertices in
a regular container,
a copy of the container is first converted
to the \emph{loose representation}, in which
the 2-bit strings corresponding to the $q$
color values are simply concatenated to
form a string of $2 q$ bits.
Since $\log_2 q\le\lambda$,
this can be done in
$O(\lambda)$ time with the algorithm of
Lemma~\ref{lem:base} below, used
with $c=3$, $d=4$ and $s=q$.
The algorithm is a word-parallel version
(i.e., essentially independent computations
take place simultaneously in different regions
of a word)
of a simple divide-and-conquer procedure.

\begin{lemma}[{\normalfont{\cite{Hag18},
Lemma 3.3 with $f=2$ and $p=1$}}\bf]
\label{lem:base}
Given integers $c$, $d$ and $s$
with $2\le c,d\le 4$, $s\ge 1$ and $s=O(w)$
and an integer of the form
$\sum_{j=0}^{s-1}a_{j}c^j$,
where
$0\le a_{j}<\min\{c,d\}$
for $j=0,\ldots,s-1$,
the integer
$\sum_{j=0}^{s-1}a_{j}d^j$
can be computed in
$O(\log(s+1))$ time.
\end{lemma}

Conversely, using the lemma instead with $c=4$
and $d=3$, we can convert from the loose to the
regular representation, again in
$O(\lambda)$ time.
Once a container is in the loose representation,
we can locate the first (smallest) gray vertex
in the container
in constant time with the algorithm of
part~(a) of the following lemma that, again,
draws heavily on word-parallel techniques.
Here we use the lemma with $m=q$ and $f=2$.

\begin{lemma}[\normalfont{\cite{HagK16},
Lemma 3.2}\bf]
\label{lem:word}
Let $m$ and $f$ be given integers
with $1\le m,f<2^w$ and suppose that a
sequence $A=(a_1,\ldots,a_m)$ with
$a_i\in\{0,\ldots,2^f-1\}$ for $i=1,\ldots,m$
is given
in the form of the $(m f)$-bit binary representation
of the integer
$\sum_{i=0}^{m-1} 2^{i f}a_{i+1}$.
Then the following holds:
\begin{itemizewithlabel}
\item[(a)]
Let $I_0=\{i\in\TbbbN:1\le i\le m$ and $a_i=0\}$.
Then, in $O(1+{{m f}/w})$ time,
we can test whether $I_0=\emptyset$ and, if not,
compute $\min I_0$.
\item[(b)]
If $m<2^f$ and
an additional integer $k\in\{0,\ldots,2^f-1\}$
is given, then $\Tvn{rank}(k,A)=
|\{i:1\le i\le m$ and $k\ge a_i\}|$
can be computed
in $O(1+{{m f}/w})$ time.
\end{itemizewithlabel}
\end{lemma}

Subsequently, if we remember the last grey vertex
enumerated, we can shift out that vertex and all
vertices preceding it before applying the same
algorithm.
This enables us to iterate over the set $V\Tsub g$
of gray vertices in the container
in $O(|V\Tsub g|+1)$ time.
The colors of the at most $q-1$ vertices left over
from the division into segments are kept
permanently in
what corresponds to the loose representation.
This uses $O(\log n)$ bits, and it will be
obvious how to adapt the various operations
to take these vertices and their colors into account,
for which reason we shall ignore them in
the following.

If a container is a slave
(Fig.~\ref{fig:rep}(d)),
we require the number $n\Tsub g$ of gray vertices in the
container to be bounded by $\lambda-1$,
and we store its gray vertices separately
in a \emph{gray list}.
The gray list takes the form of the integer $n\Tsub g$,
stored (somewhat wastefully) in $\lambda$ bits,
followed by a sorted sequence of
$n\Tsub g$ integers, each represented in
$\lambda$ bits, that indicate the
positions of the gray vertices within the container.
By the assumption $\lambda^2\le\Lambda$,
the gray list fits within $\Lambda$ bits.
Because of the availability of
the gray list,
we can store the remaining
vertex colors in a
\emph{black-and-white vector} of just $q$ bits
by dropping the most significant
bit, which normally allows us to distinguish between
the colors gray and black, from all $q$
2-bit color values.
Since $3^2\ge 2^3$ and therefore
$q\log_2 3\ge 15\Lambda$, this leaves at least
$15\Lambda-\Lambda-q=4\Lambda$ bits, which
are used to hold the $3\Lambda$ bits relocated
from the master and a pointer to the master.
We call this representation the
\emph{compact representation}.
A container may be \emph{compact},
i.e., in the compact representation,
without being a slave
(Fig.~\ref{fig:rep}(c)).
Then, instead of the data relocated from a
master, it stores two iteration pointers
and a null pointer.

Using the algorithm of Lemma~\ref{lem:word}(b)
with $m=n\Tsub g$ and $f=\lambda$,
we can test in constant time whether a vertex
located in a compact container is gray
by checking whether its number within the
container occurs in the gray list of the container.
If not, we can subsequently determine
the color of the vertex in constant
time from the black-and-white vector.
Similarly, we can change the color of a
given vertex in constant time.
This may involve creating a gap for the
new vertex in the gray list
or, conversely, closing such a gap, which is
easily accomplished
with a constant number of bitwise
Boolean and shift operations.
It is also easy to see that we can iterate over
the $n\Tsub g$ gray vertices in $O(n\Tsub g+1)$ time.

If a color change increases the number
$n\Tsub g$ of gray
vertices in a compact container to $\lambda$, the
container must be converted to the regular representation.
For this it will be convenient if the
black-and-white vector
stores the $q$ least
significant bits of the vertex colors not
in their natural order, but in the shuffled
order obtained by placing the first half of the
bits, in the natural order,
in the odd-numbered positions of the
black-and-white vector and the
last half in the even-numbered positions.
With this convention, we can still read and
update vertex colors in constant time.
We can also unshuffle the black-and-white vector
in constant time,
creating 1-bit gaps for the most significant bits,
by separating the bits in the odd-numbered
positions from those in the even-numbered
positions and concatenating the two sequences.
Subsequently each most significant bit
is set to be the complement of its
corresponding least significant bit
to represent
the colors white and black
according to the loose representation.
Going through the gray list, we can then
introduce the gray colors one by one.
Thus we can
convert from the compact
to the loose and from there to the regular
representation in
$O(n\Tsub g+\lambda)=O(\lambda)$ time.
Conversely, if a container in the loose
representation has fewer than $\lambda$ gray
vertices, it can be converted to
the compact representation in $O(\lambda)$ time.

\subsection{The In-Place Chain Technique for Containers}

The overall organization of the containers
follows the in-place chain
technique \cite{KatG17,Hag18,KamS18}.
By means of an integer $\Tmu\in\{0,\ldots,N\}$
equal to the number of compact containers,
the sequence $D_1,\ldots,D_N$ of containers
is dynamically divided into a \emph{left part},
consisting of $D_1,\ldots,D_\Tmu$, and a
\emph{right part},
$D_{\Tmu+1},\ldots,D_N$.
A regular container is a master if and only
if it belongs to the left part, and a
compact container is a slave if and only
if it belongs to the right part.
Thus the two representations shown on the
left in Fig.~\ref{fig:rep} ((a) and (c))
can occur only in the left part, while
the two representations shown on the right
can occur only in the right part.
In particular, note that every container
in the left part has iteration pointers.

Call a container \emph{gray-free} if no
vertex located in the container is gray.
The iteration pointers are used to join
all containers in the left part, with
the exception of the gray-free compact
containers, into a doubly-linked
\emph{iteration list} whose first and
last elements are stored in $O(\log n)$ bits
outside of the containers.

When an update of a vertex color causes a
container $D_i$ to switch from the compact
to the regular representation, $\Tmu$
decreases by~1, say from $\Tmu_0$ to $\Tmu_0-1$.
If $i=\Tmu_0$, $D_i$ belongs to the left part
before the switch and to the right part after
the switch, i.e., in terms of Fig.~\ref{fig:rep},
the switch is from (c) to~(b).
If $i\not=\Tmu_0$, the switch is more complicated,
in that it involves other containers.
If $i<\Tmu_0$ (Fig.~1, (c) to~(a)),
$D_i$ becomes a master,
whereas if $i>\Tmu_0$
(Fig.~1, (d) to~(b)),
$D_i$ stops being a slave.
In both cases there now is a master $D\Tsub m$
without a slave, a situation that must be remedied.
However, $D_{\Tmu_0}$ also switches,
namely either from (a) to~(b)
or from (c) to (d).
In the case ``(a) to (b)'' $D_{\Tmu_0}$
stops being a master, and its former slave can
become the slave of~$D\Tsub m$.
In the case ``(c) to (d)'' $D_{\Tmu_0}$
becomes a slave and can serve as the slave of~$D\Tsub m$.
Thus in all cases masters and slaves can again
be matched up appropriately.
Altogether, the operation involves changing some
pointers and moving some relocated data in at
most four containers.
After the conversion of $D_i$,
this takes constant time.

In some circumstances that still need to be
specified, a container $D_i$ may switch from
the regular to the compact representation,
which causes $\Tmu$ to increase by~1, say
from $\Tmu_0$ to $\Tmu_0+1$.
We can handle this situation similarly as above.
If $i=\Tmu_0+1$, the switch is from (b) to~(c)
in Fig.~\ref{fig:rep}, and nothing more
must be done.
Otherwise, whether the switch is from (a) to~(c)
or from (b) to~(d), there will be a slave
without a master.
Simultaneously $D_{\Tmu_0+1}$ switches either
from (b) to (a) (it becomes the
needed master) or from (d) to~(c)
(it stops being a slave, and its former
master takes on the new slave).
Again, after the conversion of $D_i$,
the operation can
happen in constant time.

\section{BFS Algorithms}
\label{sec:algorithms}

\subsection{The Basic Algorithm}

To execute the first line
of the abstract 3-color BFS algorithm
with the vertex-color data structure developed
in the previous section, we initialize
the data structure as follows:
All vertices are white,
all containers are compact, but not slaves,
all have empty gray lists, the
iteration list is empty, and $\Tmu=N$.

It was already described how to read and
update vertex colors.
If a color change causes a compact container $D$ in the
left part to become gray-free, $D$
is shunted out of the iteration list.
Conversely, if a compact container in the left
part stops being gray-free, it is inserted
at the end of the iteration list.
The case in which a container enters or leaves
the left part because of a change in $\Tmu$
is handled analogously.
All of this can happen in constant time.
The only exception is if a color change
forces a container to switch from the
compact to the regular representation,
which takes $O(\lambda)$ time.

Recall that the abstract 3-color BFS algorithm
alternates between \emph{exploration rounds},
in which it iterates over the gray vertices
and colors some of their white neighbors gray,
and \emph{consolidation rounds}, in which
it iterates over the gray vertices and colors
some of them black.
Each iteration is realized by iterating over
two lists of containers:
The explicitly maintained iteration list
and the implicit \emph{right list}, which
consists of the containers $D_N,D_{N-1},\ldots,D_{\Tmu+1}$
in that order.
Each of the two iterations can be viewed as
moving a \emph{pebble} through the
relevant list.
Because the lists may change dynamically,
the following rules apply:
If a currently pebbled container $D$ is deleted from
its list, the pebble is first moved to
the successor of $D$ in the relevant list.
If a pebble reaches the end of its list,
it waits there for new containers that may
be inserted at the end of the list.
One of the at most two pebbled containers
is the \emph{current container} $D\Tsub c$, whose
gray vertices are enumerated as explained earlier.
If $D\Tsub c$ is regular, this involves
first converting it to the loose representation.
Once all gray vertices in $D\Tsub c$
have been enumerated, $D\Tsub c$ stops being the
current container.
If it is in the loose representation,
we attempt to convert it to the compact representation.
If this fails because $D\Tsub c$ contains more
than $\lambda-1$ gray vertices, we instead
convert it to the regular representation.
Then the pebble on $D\Tsub c$ is moved to
the list successor of $D\Tsub c$, and one of the at most
two containers that are now pebbled is chosen to be
the new current container.
The iteration ends when
both pebbles are at the end of their
respective lists.

Since every container that is not gray-free
belongs either to the iteration list or
to the right list,
it is clear that each round
enumerates all vertices that are gray at the
beginning of the round (and maybe some that
become gray in the course of the round).
A container and its gray vertices may be
enumerated twice, namely once as part of
the iteration list and once as part of the right list.
The BFS algorithm can tolerate this,
and no vertex is enumerated more than twice
within one round because $\Tmu$ moves in only
one direction within the round.
A vertex can be gray for at most (part of)
four consecutive rounds, so the total number
of vertex enumerations is $O(n)$.
Therefore the total time spent on enumeration
is $O(n)$, except possibly for the following
two contributions to the running time:
(1) Containers that are enumerated but turn
out to be gray-free;
(2) Conversions of containers between
different representations.
As for~(1), every container concerned
is regular or on the right side, i.e., the
number of such containers is bounded by $2(N-\Tmu)$.
Since the iteration converts all
$N-\Tmu$ regular containers
to the loose representation, the contribution
of~(1) is dominated by that of~(2).
And as for~(2), since the number of other
conversions is within a constant factor of
the number of conversions to the regular
representation, it suffices to bound the latter
by~$O({n/\lambda})$.
But before the first conversion of a container
$D$ to the regular representation, $\lambda$
vertices located in $D$ must
become gray, and between two successive
such conversions at least $\lambda$
vertices in $D$ either change color or are enumerated.
Since the number of color changes and of
vertex enumerations
is $O(n)$, the bound follows.

\begin{theorem}
\label{thm:main}
The BFS problem can
be solved on directed or undirected graphs
with $n$ vertices and $m$ edges in
$O(n+m)$ time with
$n\log_2 3+O((\log n)^2)$
bits of working memory.
\end{theorem}

\subsection{A Time-Space Tradeoff}

In order to derive a time-space tradeoff
from Theorem~\ref{thm:main}, we must take
a slightly closer look at the data structure
of Dodis et al.~\cite{DodPT10} behind
Lemma~\ref{lem:dodis}.
For a certain set $S$ whose elements can
be represented in $O(\log n)$ bits,
a certain function $g:S\to S$ that can
be evaluated in constant time and a
certain start value $x_0\in S$
that can be computed in constant time,
the preprocessing of the data structure
serves to compute and store
a table $Y$ of
$x_0=g^{(0)}(x_0),g^{(1)}(x_0),g^{(2)}(x_0),
\ldots,g^{(\Tfloor{\log_2 N})}(x_0)$,
where $g^{(j)}$, for integer $j\ge 0$, denotes $j$-fold
repeated application of~$g$.
In addition, we need the powers
$3^0,3^1,\ldots,3^{q-1}$,
which are also assumed to be stored in~$Y$.
If we carry out the preprocessing but store $g^{(j)}(x_0)$
and $3^j$
only for those values of $j$ that are divisible by $t$
for some given integer $t\ge 1$,
the shortened table $Y'$ occupies only
$O(\Tceil{{{(\log n)}/t}}\log n)=
O({{(\log n)^2}/t}+\log n)$ bits,
and the rest of the BFS algorithm works
with $O(\log n)$ bits.
Whenever the data structure
of Section~\ref{sec:rep}
is called upon to carry out an operation,
it needs a constant number of entries of~$Y$,
which can be reconstructed from those
in $Y'$ in $O(t)$ time.
This causes a slowdown of $O(t)$ compared to
an algorithm that has the full table~$Y$
at its disposal.
Thus Theorem~\ref{thm:main}
generalizes as follows:

\begin{theorem}
\label{thm:tradeoff}
For every given $t\ge 1$, the BFS problem can
be solved on directed or undirected graphs
with $n$ vertices and $m$ edges in
$O((n+m)t)$ time with
$n\log_2 3+O({{(\log n)^2}/t}+\log n)$
bits of working memory.
\end{theorem}

\subsection{BFS with $n\log_2 3+O(\log n)$ Bits}

Suppose now that we are allowed only
$O(\log n)$ extra bits.
Then, with notation as in the previous subsection,
we can no longer afford to store the table $Y$ of 
$x_0,g(x_0),g^{(2)}(x_0),\ldots,g^{(\Tfloor{\log_2 N})}(x_0)$
and $3^0,3^1,\ldots,3^{q-1}$.
Instead we store only the two $O(\log n)$-bit
quantities $x_0$ and 3 and compute $g^{(j)}(x_0)$
and $3^j$ from them as needed.
Concerning the latter, $3^j$ can be computed
in $O(\log q)=O(\lambda)$ time for arbitrary
$j\in\{0,\ldots,q-1\}$ by a well-known method
based on repeated squaring.

When the data structure of
Dodis et al.~\cite{DodPT10} is used to represent
an array $A$ with index set $\{1,\ldots,N\}$,
$A[j]$, for $j=1,\ldots,N$,
is associated with the node~$j$
in a complete $N$-node binary tree $T$ whose nodes
are numbered $1,\ldots,N$ in the manner
of Heapsort, i.e.,
the root is~1 and
the parent of every nonroot node~$j$
is $\Tfloor{j/2}$.
Suppose that a node
$j\in\{1,\ldots,N\}$ is of height $h$ in~$T$.
Then we can access (read or update) $A[j]$
in constant time after computing
$g^{(h)}(x_0)$ from $x_0$, which takes $O(h+1)$ time.
In the worst case $h=\Theta(\log n)$, so we can
access $A$ with a slowdown of $O(\log n)$
relative to an algorithm with access to
the full table~$Y$.
This leads to the result of Theorem~\ref{thm:tradeoff}
for $t=\log n$, i.e.,
$O((n+m)\log n)$ time and $O(\log n)$ extra bits.
However, for most $j$ the height $h$
is much smaller than $\log_2 n$, which hints
at a possible improvement.

For $i=1,\ldots,n$, let $d_i$ be the (total) degree
of the vertex $i$ in the input graph.
It is easy to see that the number of accesses
to the color of $i$ in the course
of the execution of the BFS
algorithm is $O(d_i+1)$.
The color of $i$ is located in the container
$D_j$, where $j=\Tceil{{i/q}}$,
or in a slave $D_{j'}$ with $j'>j$,
and $D_j$ is in fact
a big digit stored in $A[j]$, where $A$ is
the array maintained with the data structure
of Dodis et al.~\cite{DodPT10}.
The depth of the node $j$ in the
corresponding binary tree $T$ is exactly
$\Tfloor{\log_2 j}=\Tfloor{\log_2\Tceil{i/q}}$,
and it is not difficult
to see that its height is at most
$\Tfloor{\log_2\Tceil{n/q}}-\Tfloor{\log_2\Tceil{i/q}}
\le\log_2({n/i})+2$.
Therefore the running time of the complete
BFS algorithm is
$O((n+m)\log\log n+\sum_{i=1}^n d_i\log({{2 n}/i}))$.

If $d_1\le d_2\le\cdots\le d_n$,
$\sum_{i=1}^n d_i\log_2({{2 n}/i})\le
({1/n})(\sum_{i=1}^n d_i)(\sum_{i=1}^n\log_2({{2 n}/i}))
=O(m)$.
Thus if the vertex degrees form a nondecreasing
sequence, the running time is
$O((n+m)\log\log n)$.
Since $\log_2({{2 n}/i})\le 1+r\log_2\log_2 n$
if $i\ge{n/{(\log_2 n)^r}}$ for some $r\ge 1$,
the same is true if
$\sum_{i=1}^{\Tfloor{{n/{(\log_2 n)^r}}}} d_i=
O({{m\log\log n}/{\log n}})$
for some fixed $r\ge 1$.
Informally, the latter condition is satisfied
if $G$ is approximately regular.
In particular, it is satisfied if the ratio of
the maximum degree in $G$ to the average degree is
(at most) polylogarithmic in~$n$.

\bibliography{all}

\end{document}